\documentclass[aps,prl,twocolumn,superscriptaddress,showpacs]{revtex4}

\usepackage{graphicx}

\begin{document}

\title{Entanglement and Quantum Phase Transition Revisited}

\author{Min-Fong Yang}
\affiliation{Department of Physics, Tunghai University, Taichung,
Taiwan}

\date{\today}

\begin{abstract}
We show that, for an exactly solvable quantum spin model, a
discontinuity in the first derivative of the ground state
concurrence appears in the absence of quantum phase transition. It
is opposed to the popular belief that the non-analyticity property
of entanglement (ground state concurrence) can be used to
determine quantum phase transitions. We further point out that the
analyticity property of the ground state concurrence in general
can be more intricate than that of the ground state energy. Thus
there is no one-to-one correspondence between quantum phase
transitions and the non-analyticity property of the concurrence.
Moreover, we show that the von Neumann entropy, as another measure
of entanglement, can not reveal quantum phase transition in the
present model. Therefore, in order to link with quantum phase
transitions, some other measures of entanglement are needed.
\end{abstract}


\maketitle


Quantum entanglement, as one of the most fascinating feature of
quantum theory, has attracted much attention over the past decade,
mostly because its nonlocal connotation~\cite{ABinstein35} is
regarded as a valuable resource in quantum communication and
information processing~\cite{MANielsenb}. Recently a great deal of
effort has been devoted to the understanding of the connection
between quantum entanglement and quantum phase transitions
(QPTs)~\cite{Osterloh:02,Osborne:02,Gu:03,Syl:03,JVidal:04a,
Vidal:03,Vidal:04,Verstraete:04a,Wei:04,Bose:02,JVidal:04b,
Lambert:04,Gu:04}. Quantum phase transitions~\cite{Sachdev:book}
are transitions between qualitatively distinct phases of quantum
many-body systems, driven by quantum fluctuations.  In view of the
connection between entanglement and quantum
correlations~\cite{Glaser:03}, one anticipates that entanglement
will furnish a dramatic signature of the quantum critical point.
People hope that, by employing quantum entanglement, the global
picture of the quantum many-body systems could be diagnosed, and
one may obtain fresh insight into the quantum many-body problem.
Hence, in addition to its intrinsic relevance with quantum
information applications, entanglement may also play an
interesting role in the context of statistical mechanics.

The aforementioned studies are based on the analysis of particular
many-body models. Recently a theorem of the relation between QPTs
and bipartite entanglement is proposed~\cite{WuSarandyLidar04}.
The authors conclude that, under certain conditions, a
discontinuity in or a divergence of the ground state concurrence
[the first derivative of the ground state concurrence] is both
necessary and sufficient to signal a first-order QPT (1QPT)
[second-order QPT (2QPT)]. Most of the previous investigations for
specific models support their conclusion. This may strengthen the
belief that one can determine QPTs by using quantum entanglement.

In this paper, the entanglement properties (the ground state
concurrence and the von Neumann entropy) are calculated for an
exactly solvable quantum spin model~\cite{LWC}. Contrary to
conventional wisdom, we find that there exists a discontinuity in
the first derivative of the concurrence, {\it at which there is no
quantum critical point}. In fact, similar result had already been
discovered in Ref.~\cite{JVidal:04a} for a quantum spin model on a
simplex in a magnetic field. Here we give general arguments to
show why the analyticity property of the concurrence is more
intricate than that of the ground state energy. Thus there is no
one-to-one correspondence between QPTs and the non-analyticity
property of the concurrence. That is, it is not always possible to
infer the existence of QPTs from concurrence. Furthermore, for the
one-dimensional $XXZ$ model at the critical point of the isotropic
ferromagnetic case, it is found that the first derivative of the
concurrence is discontinuous~\cite{Syl:03}. However, it is a 1QPT,
instead of 2QPT. The reason why the non-analyticity of the ground
state energy of the $XXZ$ spin chain does not faithfully reflect
that of concurrence is explained. From
Refs.~\cite{Syl:03,JVidal:04a} and our result, it is clear that
QPTs in general can not be distinctly characterized through the
analysis of the analyticity properties of concurrence. Moreover,
we show that, for the model considered in this paper, the von
Neumann entropy remains constant even crossing the critical point.
That is, the von Neumann entropy can not always detect QPTs.
Therefore, in order to have close connection with QPTs, some other
measures of entanglement are needed.


The exactly solvable quantum spin model considered here is the
isotropic spin-$\frac{1}{2}$ $XY$ (or spin-$\frac{1}{2}$ $XX$ )
chain with three-spin interactions~\cite{LWC},
\begin{eqnarray}\label{eq:model}
H= -\sum_{i=1}^{N} &&\left[ \sigma_i^x \sigma_{i+1}^x + \sigma_i^y
\sigma_{i+1}^y \right. \nonumber  \\
&& \left. + \frac{\lambda}{2} \left(\sigma_{i-1}^x \sigma_i^z
\sigma_{i+1}^y - \sigma_{i-1}^y \sigma_i^z \sigma_{i+1}^x \right)
\right] ,
\end{eqnarray}
where $N$ is the number of sites, $\sigma_i^\alpha$ ($\alpha
=x,y,z$) are the Pauli matrices, and $\lambda$ is a dimensionless
parameter characterizing the three-spin interaction strength (in
unit of the nearest-neighbor exchange coupling). Periodic boundary
condition $\sigma_{N+1}=\sigma_1$ is assumed. This model can be
solved by using the Jordan-Wigner transformation~\cite{JW,note1},
and all physical quantities can in principle be calculated
exactly. It is shown that the three-spin interaction can lead to a
2QPT at $\lambda_c =1$~\cite{LWC}.

Here we consider the entanglement between two specific spins in
the ground state of a quantum system. The state of two spins $i$
and $j$ in the ground state of a quantum system is described in
terms of the two-particle reduced density matrix $\rho_{ij}$
obtained by tracing over other spins.

The structure of $\rho_{ij}$ follows from the symmetry properties
of the Hamiltonian. The Hamiltonian in Eq.~(\ref{eq:model}) is
real and it has the following two symmetries. One is a global
$U(1)$-rotation symmetry about the spin-$z$ axis,
another is a $Z_2$ symmetry of a global $\pi$-rotation about the
spin-$x$ axis~\cite{note2}.
These symmetries guarantee that $\rho_{ij}$ has the
form~\cite{O'Connor}
\begin{equation}
\rho_{ij}=\left(
\begin{array}{llll}
u_{ij} & 0 & 0 & 0 \\
0& w_{ij} & z_{ij} &0  \\
0& z_{ij} & w_{ij} & 0 \\
0& 0 & 0 & u_{ij}
\end{array}
\right)   \label{eq:rho12}
\end{equation}
in the standard basis $ \{ |\uparrow \uparrow \rangle,|\uparrow
\downarrow \rangle,|\downarrow \uparrow \rangle, |\downarrow
\downarrow \rangle \}$. Wang and Zanardi~\cite{XWang02} have shown
that the matrix elements of $\rho_{ij}$ can be expressed in terms
of the various correlation functions $\langle \sigma_i^\alpha
\sigma_j^\beta \rangle$ ($\alpha, \beta =x,y,z$):
\begin{eqnarray}\label{eq:uwz}
u_{ij}&=&\frac{1}{4} \left( 1+\langle \sigma_i^z \sigma_j^z \rangle \right), \nonumber \\
w_{ij}&=&\frac{1}{4} \left( 1-\langle \sigma_i^z \sigma_j^z \rangle \right), \\
z_{ij}&=&\frac{1}{4}\left(\langle \sigma_i^x \sigma_j^x \rangle +
\langle \sigma_i^y \sigma_j^y \rangle \right). \nonumber
\end{eqnarray}
Note that $u_{ij}, w_{ij}\geq 0$ because of the inequality
$|\langle \sigma_i^z \sigma_j^z \rangle|\le 1$, which is a special
case of the Schwarz inequality $|\langle A^\dag B\rangle|\le
\sqrt{\langle A^\dag A\rangle} \sqrt{\langle B^\dag B\rangle}$ for
$A=I$ ($I$ is the identity operator) and $B=\sigma_i^z
\sigma_j^z$.

From $\rho_{ij}$, the ground state concurrence~\cite{Wootters:98}
quantifying the entanglement is readily obtained
as~\cite{O'Connor,XWang02}
\begin{eqnarray}
C_{i,j}&=&2\max \left\{ 0,|z_{ij}|-u_{ij} \right\}  \nonumber\\
       &=&\frac{1}{2}\max \left\{ 0,|\langle\sigma_i^x \sigma_j^x \rangle
+ \langle \sigma_i^y \sigma_j^y \rangle | - \langle \sigma^z_i
\sigma^z_j \rangle -1  \right\}.        \label{eq:concurrence1}
\end{eqnarray}
Because the entanglement between a pair of adjacent spins is
expected to be dominant compared with a pair of
non-nearest-neighbor spins, we focus on the nearest-neighbor case
in the following discussions.

Using the method adopted by Lieb, Schultz and Mattis~\cite{Lieb},
one can calculate the spin-spin correlation functions
exactly~\cite{LWC},
\begin{eqnarray}\label{eq:correlation}
\langle \sigma^{x}_i \sigma^{x}_{i+1} \rangle &=& \langle
\sigma^{y}_i \sigma^{y}_{i+1}\rangle = G , \nonumber\\
\langle \sigma^{z}_i \sigma^{z}_{i+1} \rangle &=& -G^2,
\end{eqnarray}
with
\begin{eqnarray}\label{eq:G}
  G&=&
  \left\{
  \begin{array}{lc}
    \displaystyle{\frac{2}{\pi}} & \displaystyle{~~~\lambda < 1},
    \smallskip\\
    \displaystyle{\frac{2}{\pi\lambda}}
    & \displaystyle{~~~\lambda \geq 1}.
  \end{array}
    \right.
\end{eqnarray}
We find that, for the nearest-neighbor cases, two correlation
functions $\langle \sigma^{x}_i \sigma^{x}_{i+1} \rangle$ and
$\langle \sigma^{z}_i \sigma^{z}_{i+1} \rangle$ are dependent, and
the latter can be written in terms of the former. Thus the
nearest-neighbor concurrence is only determined by a single
correlation function $G$. By substituting the results of the
correlation functions into Eq.~(\ref{eq:concurrence1}), the exact
expression of the concurrence between a pair of adjacent spins
becomes
\begin{equation}\label{eq:concurrence2}
C_{i,i+1}=\frac{1}{2}\max \left\{ 0, (G+1)^2 -2 \right\} .
\end{equation}
The dependence of $C_{i,i+1}$ on $\lambda$ is plotted in Fig.~1.
We see that, both at $\lambda = \lambda_c=1$ and $\lambda
=\lambda_0 = 2/(\sqrt{2}-1)\pi \approx 1.5369$, the first
derivative of the concurrence $\partial C_{i,i+1} /\partial
\lambda$ shows discontinuities, while $C_{i,i+1}$ is continuous.
The discontinuity in $\partial C_{i,i+1} /\partial \lambda$ at
$\lambda = \lambda_c =1$ do indicate the 2QPT of the present
model, consistent with the proposal in
Ref.~\cite{WuSarandyLidar04}. However, an unexpected discontinuity
in $\partial C_{i,i+1} /\partial \lambda$ occurs at $\lambda
=\lambda_0$, which is {\it not} a critical point! The origin of
non-analyticity in the concurrence at $\lambda =\lambda_0$ comes
from the requirement that the concurrence should be non-negative,
but not from the non-analyticity of $\rho_{ij}$. Therefore, the
discontinuity in $\partial C_{i,i+1} /\partial \lambda$ 
needs not always indicate the existence of non-analyticity in the
ground state energy and show any QPT. We note that the possibility
of such an unanticipated discontinuity in $\partial C_{i,i+1}
/\partial \lambda$ is not addressed by the theorem of
Ref.~\cite{WuSarandyLidar04}.
\begin{figure}[ht]
\centering {\includegraphics[width=2.5in]{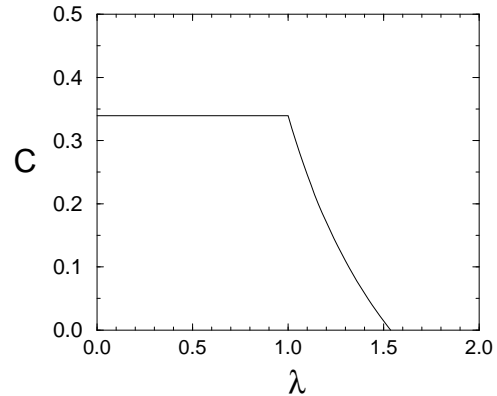}} %
\caption{The ground state concurrence of the nearest-neighbor
spins $C_{i,i+1}$ as a function of $\lambda$ for the $XX$ chain
with three-spin interactions in Eq.~(\ref{eq:model}). }
\end{figure}

In a recent work studying the one-dimensional extended Hubbard
model~\cite{Gu:04}, the authors show that QPTs can be identified
at places where the von Neumann entropy is extremum or its
derivative is singular. The von Neumann entropy, another measure
of entanglement, is defined as $S \equiv -{\rm tr}(\rho_j \log_2
\rho_j)$, where $\rho_j$ is the one-particle reduced density
matrix obtained by tracing over all sites except the $j$-th site,
and therefore $\rho_j = {\rm tr}_i(\rho_{ij})$ (${\rm tr}_i$
stands for tracing over the $i$-th site). One may wonder if this
measure of entanglement will still work for the present model. By
using Eqs.~(\ref{eq:rho12}) and (\ref{eq:uwz}) and after tracing
over the $i$-th site for $\rho_{ij}$, one obtains $\rho_j
=\frac{1}{2} I$,
and the von Neumann entropy $S=1$
for all $\lambda$. The von Neumann entropy thus fails to detect
the QPT of the present model. It is because the non-analyticity in
the matrix elements of $\rho_{ij}$ cancel each other by taking
trace over the $i$-th site. Hence the von Neumann entropy is not
always useful to detect QPT.

We note that the present example is not the only exception for the
anticipation that the non-analyticity property of concurrence can
be used to determine QPT. In general, the
concurrence~\cite{Wootters:98} is defined by $C_{ij}=\max \{ 0,
\tilde{C}_{ij}  \}$ with $\tilde{C}_{ij} \equiv \gamma_1 -
\gamma_2 - \gamma_3 - \gamma_4$. Here $\gamma_\alpha$ are the four
eigenvalues , in descending order, of the matrix $R_{ij} \equiv
\sqrt{\sqrt{\rho_{ij}} \;(\sigma^y_i \otimes \sigma^y_j
{\rho}^*_{ij} \sigma^y_i \otimes \sigma^y_j) \sqrt{\rho_{ij}}}$
with $\rho^{\ast}_{ij}$ being the complex conjugate of the reduced
density matrix $\rho_{ij}$ in the standard basis. We emphasize
that, even for the models such that $\tilde{C}_{ij}$ is always
non-negative and {\it the matrix elements of $\rho_{ij}$ change
smoothly} as the physical parameter, say $\lambda$, is varied, the
concurrence (in this case $C_{ij}=\tilde{C}_{ij}$) can still show
a cusp-like singularity at some $\lambda=\lambda_0$. For example,
this will happen (cf.
Eq.~(7) of Ref.~\cite{Glaser:03}) if 
at least one of the $\gamma_{\alpha}$ takes the form of $|A-B|$
(where $A$ and $B$ denote two functions of $\lambda$) and $A-B$
changes sign at $\lambda=\lambda_0$. Therefore, the concurrence
can be non-analytic, but it again does not correspond to a QPT.
That is, although the eigenvalues $\gamma_{\alpha}$ are
algebraically related to the matrix elements of $\rho_{ij}$,
$\tilde{C}_{ij}$ may still have {\it different} analyticity
properties from $\rho_{ij}$. Reminding that $\max \{ 0,
\tilde{C}_{ij} \} = |\tilde{C}_{ij}|/2 + \tilde{C}_{ij}/2$, it is
clear that the two possibilities of the unexpected discontinuities
in $\partial C_{i,i+1} /\partial \lambda$ discussed above all
originate from the non-analyticity of the absolute-value function.
As mentioned before, an unexpected discontinuity in $\partial
C_{i,i+1} /\partial \lambda$, which does not indicate a QPT, had
already been discovered in Ref.~\cite{JVidal:04a}. We believe that
the non-analyticity in the concurrence in that case may be due to
the reason explained above.

Even though the discontinuity in the first derivative of the
concurrence does indicate a QPT, it may not be 2QPT. An example is
the one-dimensional $XXZ$ model,
\begin{equation}\label{eq:XXZ}
H_{XXZ}=\sum_{i=1}^N [\sigma_i^x \sigma_{i+1}^x+\sigma_i^y
\sigma_{i+1}^y+\Delta \sigma_i^z \sigma_{i+1}^z].
\end{equation}
It is shown that, at the critical point $\Delta=-1$ (corresponding
to the ferromagnetic Heisenberg model), $\partial C_{i,i+1}
/\partial \Delta$ is discontinuous, while $C_{i,i+1}$ is
continuous and $C_{i,i+1}|_{\Delta=-1} =0$~\cite{Syl:03}. However,
it is a 1QPT (see below), instead of 2QPT. The reason why the
non-analyticity of the concurrence of the $XXZ$ spin chain does
not faithfully correspond to that of the ground state energy is
explained below.

We first show the relations among the ground state energy (and its
first derivative), the matrix elements of the reduced density
matrix, and the concurrence for the $XXZ$ spin chain. The
concurrence of the $XXZ$ spin chain has the same expression as
Eq.~(\ref{eq:concurrence1})~\cite{XWang02,Glaser:03,Gu:03,note4}.
Due to the translational invariance, the ground state energy per
site for the $XXZ$ spin chain can be written as $\mathcal{E}
=\langle\sigma_i^x \sigma_{i+1}^x \rangle + \langle\sigma_i^y
\sigma_{i+1}^y \rangle + \Delta \langle\sigma_i^z \sigma_{i+1}^z
\rangle$. Employing the Hellmann-Feynman theorem~\cite{HF}, one
has $\partial \mathcal{E}/\partial \Delta= \langle\sigma_i^z
\sigma_{i+1}^z \rangle$. Thus, for the nearest-neighbor spins, two
matrix elements $u_{i,i+1}$ and $z_{i,i+1}$ of the reduced density
matrix can be written as
\begin{eqnarray}
u_{i,i+1}&=&\frac{1}{4} \left( 1 + \frac{\partial
\mathcal{E}}{\partial \Delta} \right), \nonumber \\
z_{i,i+1}&=&\frac{1}{4}\left( \mathcal{E} - \Delta \frac{\partial
\mathcal{E}}{\partial \Delta} \right),
\end{eqnarray}
and the nearest-neighbor concurrence becomes $C_{i,i+1}=\max \{0,
\tilde{C}_{i,i+1} \}$ with
\begin{equation}
\tilde{C}_{i,i+1}=-\frac{1}{2}\left[(\mathcal{E} +1) + (1-\Delta)
\frac{\partial \mathcal{E}}{\partial \Delta} \right] .
\end{equation}
Here we use the fact that $\langle\sigma_i^x \sigma_{i+1}^x
\rangle + \langle\sigma_i^y \sigma_{i+1}^y \rangle \leq 0$. This
inequality is satisfied for the $XXZ$ spin chain in
Eq.~(\ref{eq:XXZ}), because the ground state wavefunction obeys
the Marshall-Peierls sign rule~\cite{Marshall}.

The $XXZ$ spin chain is an exactly solvable model, and the
expression of $\mathcal{E}$ can be found in Ref.~\cite{Yang}. For
the critical point $\Delta=-1$, one has
$\mathcal{E}|_{\Delta=-1}=-1$, $\partial \mathcal{E}/\partial
\Delta \rightarrow 0$ as $\Delta \rightarrow -1^+$ and $\partial
\mathcal{E}/\partial \Delta =1$ for $\Delta < -1$. Thus $\partial
\mathcal{E}/\partial \Delta$ is discontinuous at $\Delta=-1$,
which is a manifestation of a 1QPT. Based on these results, we
find that $\tilde{C}_{i,i+1}$ is indeed not continuous at
$\Delta=-1$, where it has a finite jump from $-1$ to 0. However,
because $C_{i,i+1}=\max \{ 0, \tilde{C}_{i,i+1} \}$, $C_{i,i+1}$
becomes continuous and equal to zero at $\Delta=-1$. That is, the
discontinuity in $\tilde{C}_{i,i+1}$ is hidden under the operation
$\max \{ 0, \dots \}$. That is the reason why the non-analyticity
of the concurrence is not faithfully induced by that of the ground
state energy. In short, the discontinuity in the first derivative
of $\mathcal{E}$ (and therefore the matrix elements of the reduced
density matrix) may not always lead to discontinuity in
$C_{i,i+1}$. Therefore, a 1QPT may be misunderstood as a 2QPT
through analyzing the non-analyticity property of concurrence.

There is another critical point of the $XXZ$ spin chain at
$\Delta=1$ (corresponding to the anti-ferromagnetic Heisenberg
model). It is found that $C_{i,i+1}$ and $\partial C_{i,i+1}
/\partial \Delta$ are both continuous at $\Delta=1$, and $\partial
C_{i,i+1} /\partial \Delta|_{\Delta=1} =0$ (or $C_{i,i+1}$ reaches
its maximum value at $\Delta=1$)~\cite{Gu:03,Syl:03}. It is
interesting to see how these results can be realized in the
present framework. At the critical point $\Delta=1$, it is shown
in Ref.~\cite{Yang} that $\mathcal{E}$ and all of its derivatives
with respect to $\Delta$ are continuous. Therefore, $C_{i,i+1}$
($=\tilde{C}_{i,i+1}$, because $\tilde{C}_{i,i+1}\geq 0$ in this
case) and $\partial C_{i,i+1} /\partial \Delta$ will not show
discontinuity at $\Delta=1$. 
Moreover, since
\begin{equation}
\frac{\partial C_{i,i+1}}{\partial \Delta}=-\frac{1}{2} (1-\Delta)
\frac{\partial^2 \mathcal{E}}{\partial \Delta^2} ,
\end{equation}
we find that $\partial C_{i,i+1} / \partial \Delta \rightarrow 0$
as $\Delta \rightarrow 1$. Thus the results in
Refs.~\cite{Gu:03,Syl:03} are reproduced.


In summary, although many examples indicate that QPTs can be
distinctly characterized through the analyticity properties of
concurrence, we stress in this paper that this viewpoint is not
true in general. Except those cases of 2QPTs indicated by the
discontinuity in $\partial C_{i,i+1} /\partial \Delta$, it is also
possible that (i) $\partial C_{i,i+1} /\partial \Delta$ is
discontinuous, but there is no QPT (Ref.~\cite{JVidal:04a} and the
present case); (ii) $\partial C_{i,i+1} /\partial \Delta$ is
discontinuous, while it is a 1QPT rather than 2QPT ($XXZ$ spin
chain at $\Delta=-1$~\cite{Syl:03}). We further point out that
QPTs can not always be diagnosed even by using the von Neumann
entropy. As far as we know, there are some other measures of
entanglement, say localizable entanglement~\cite{Verstraete:04a}
and global measure of entanglement~\cite{Wei:04}. Therefore, while
the analyticity properties of concurrence and von Neumann entropy
are not necessarily related to the existence of critical points,
other measures of entanglement may work. Thus more effort is
necessary to clarify the relationship between QPTs and
entanglement.

\begin{acknowledgments}
The author is grateful to M.-C. Chang for showing me their
manuscript before publication and for many valuable discussions. I
would also like to thank S.-J. Gu, D. Lidar, F. Verstraete, and J.
Vidal for their helpful comments on the manuscript. This work was
supported by the National Science Council of Taiwan under Contract
No.  NSC 92-2112-M-029-006.
\end{acknowledgments}



\begin{thebibliography}{150}

\bibitem{ABinstein35}
A. Einstein, B. Podolsky, and N. Rosen, Phys. Rev. {\bf 47}, 777
(1935).

\bibitem{MANielsenb}
M. A. Nielsen and I. L. Chuang, {\it Quantum Computation and
Quantum Information} (Cambridge University Press, Cambridge,
2000).


\bibitem{Osterloh:02}
{A. Osterloh, L. Amico, G. Falci, and R. Fazio}, Nature {\bf 416},
608 (2002).

\bibitem{Osborne:02}
{T. J. Osborne and M. A. Nielsen}, Phys. Rev. A {\bf 66},  032110
(2002).

\bibitem{Gu:03}
{S.-J. Gu, H.-Q. Lin, and Y.-Q. Li}, Phys. Rev. A {\bf 68}, 042330
(2003).

\bibitem{Syl:03}
{O. F. Sylju\aa sen}, Phys. Rev. A {\bf 68}, 060301(R) (2003).

\bibitem{JVidal:04a}
{J. Vidal, G. Palacios, and R. Mosseri}, Phys. Rev. A {\bf 69},
022107 (2004).

\bibitem{Vidal:03}
{G. Vidal, J. I. Latorre, E. Rico, and A. Kitaev}, Phys. Rev.
Lett. {\bf 90}, 227902  (2003).

\bibitem{Vidal:04}
{J. I. Latorre, E. Rico, and G. Vidal}, Quant. Inf. Comp. {\bf 4},
48  (2004).


\bibitem{Verstraete:04a}
{F. Verstraete, M. Popp, and J. I. Cirac}, Phys. Rev. Lett. {\bf
92}, 027901 (2004).




\bibitem{Wei:04}
T.-C. Wei, D. Das, S. Mukhopadyay, S. Vishveshwara, and P. M.
Goldbart, eprint quant-ph/0405162.

\bibitem{Bose:02}
{I. Bose and E. Chattopadhyay}, Phys. Rev. A {\bf 66},  062320
(2002).

\bibitem{JVidal:04b}
{J. Vidal, R. Mosseri, and J. Dukelsky}, Phys. Rev. A {\bf 69},
054101 (2004).

\bibitem{Lambert:04}
{N. Lambert, C. Emary, and T. Brandes}, Phys. Rev. Lett. {\bf 92},
073602 (2004).

\bibitem{Gu:04}
{S.-J. Gu, S.-S. Deng, Y.-Q. Li, and H.-Q. Lin}, {eprint
quant-ph/0405067}


\bibitem{Sachdev:book}
{S. Sachdev}, {\em {Quantum Phase Transitions}} ({Cambridge
University Press}, {Cambridge, UK}, 1999).

\bibitem{Glaser:03}
{U. Glaser, H. Buttner, and H. Fehske}, Phys. Rev. A {\bf 68},
032318 (2003).

\bibitem{WuSarandyLidar04}
L.-A. Wu, M. S. Sarandy, and D. A. Lidar, {eprint
quant-ph/0407056}.



\bibitem{LWC}
P. Lou, W.-C. Wu, and M.-C. Chang, Phys. Rev. B, (to be published)
and references therein.

\bibitem{JW}
P. Jordan and E. Wigner, Z. Physik {\bf 47}, 631 (1928).

\bibitem{note1}
After the Jordan-Wigner transformation, the Hamiltonian in
Eq.~(\ref{eq:model}) is mapped onto a free spinless fermion model,
where the three-spin interaction terms correspond to the {\it
two-body} terms of next-nearest-neighbor hopping with {\it
imagnary} hopping integrals~\cite{LWC}.

\bibitem{note2}
In the language of spinless fermions (Jordan-Wigner fermions),
these symmetries correspond to those with respect to the global
$U(1)$ phase and the particle-hole transformations.

\bibitem{O'Connor}
K. M. O'Connor and W. K. Wootters, Phys. Rev. A {\bf 63}, 052302
(2001).

\bibitem{XWang02}
X. Wang and P. Zanardi, Phys. Lett. A {\bf 301}, 1 (2002); X.
Wang, Phys. Rev. A {\bf 66}, 034302 (2002).

\bibitem{Wootters:98}
W. K. Wootters, Phys. Rev. Lett. {\bf 80}, 2245 (1998).

\bibitem{Lieb}
E. Lieb, T. Schultz, and D. Mattis, Ann. Phys. (N.Y.) {\bf 16},
407 (1961); P. Pfeuty, {\it ibid.} {\bf 57}, 79 (1970).


\bibitem{note4}
For $|\Delta|>1$, the ground state is doubly degenerate because of
the $Z_2$ symmetry of a global $\pi$-rotation about the spin-$x$
axis.  In this case, the entanglement will be calculated from the
thermal density matrix (at zero temperature) $\rho_{\rm th}={1
\over 2}\left(|+\rangle \langle + |  + |-\rangle
\langle-|\right),$ where $|+\rangle$ and $|-\rangle$ are the two
orthogonal ground states. Therefore, $\langle \sigma_i^z
\rangle=0$.


\bibitem{HF}
H. Hellmann, {\it Die Entfurhung in die Quantenchemie} (Deuticke,
Leipzig, 1937), p. 285; R. P. Feynman, Phys. Rev. {\bf 56}, 340
(1939).


\bibitem{Marshall}
W. Marshall, Proc. R. Soc. London Ser. {\bf A 232}, 48 (1955). See
also L. Capriotti, Int. J. Mod. Phys. {\bf B 15},  1799 (2001),
Sec.~2.1.

\bibitem{Yang}
C. N. Yang  and C. P. Yang, Phys. Rev. {\bf 150}, 321 (1966); {\it
ibid.} {\bf 150}, 327 (1966). 




\end{thebibliography}
\end{document}